\documentstyle[12pt]{article}
\begin{document}
\baselineskip 12pt
\textwidth    170mm
\textheight   220mm
\columnsep     38pt
\topmargin    -30pt
\oddsidemargin  12pt
\pagestyle{empty}
\newcommand{\be}{\begin{equation}}
\newcommand{\ee}{\end{equation}}
\newcommand{\bea}{\begin{eqnarray}}
\newcommand{\eea}{\end{eqnarray}}
\newcommand{\len}{\lefteqn}
\newcommand{\nn}{\nonumber}
\newcommand{\muh}{\hat\mu}
\newcommand{\dlr}{\stackrel{\leftrightarrow}{D} _\mu}
\newcommand{\vnew}{$V^{\rm{NEW}}$}
\newcommand{\vecp}{$\vec p$}
\newcommand{\dof}{{\rm d.o.f.}}
\newcommand{\prd}{Phys. Rev. \underline}
\newcommand{\pl}{Phys. Lett. \underline}
\newcommand{\prl}{Phys. Rev. Lett. \underline}
\newcommand{\np}{Nucl.Phys. \underline}
\newcommand{\vvp}{v_B\cdot v_D}
\newcommand{\dl}{\stackrel{\leftarrow}{D}}
\newcommand{\dr}{\stackrel{\rightarrow}{D}}
\newcommand{\mev}{{\rm MeV}}
\newcommand{\gev}{{\rm GeV}}
\newcommand{\calp}{{\cal P}}
\newcommand{\pinc}{\vec p \hskip 0.3em ^{inc}}
\newcommand{\pout}{\vec p \hskip 0.3em ^{out}}
\newcommand{\ptr}{\vec p \hskip 0.3em ^{tr}}
\newcommand{\pbr}{\vec p \hskip 0.3em ^{br}}
\newcommand{\no}{\noindent}
\newcommand{\ra}{\rightarrow}
\def\geq {\,\raisebox{-.6ex}{$\stackrel{>}{\sim}$}\,}
\def\tvi{\vrule height 12pt depth 6pt width 0pt}
\def\tv{\tvi\vrule}
\def \cc #1 {\kern .7em \hfill #1 \hfill \kern .7em}
\def\sq{\hbox {\rlap{$\sqcap$}$\sqcup$}}
\overfullrule=0pt
\font\boldgreek=cmmib10 \textfont9=\boldgreek
\mathchardef\mymu="0916 \def\bmu{{\fam=9 \mymu}\fam=1}
\begin{flushright}
 LPTHE Orsay 94/03\\
DAPNIA/SPP/94-23\\
hep-ph/9407406\\
 June 1994
\end{flushright}
\vskip 1 cm
\centerline{\bf DETERMINATION OF THE CP VIOLATING PHASE $\gamma$}
\par
 \centerline{\bf BY A SUM OVER COMMON DECAY MODES TO $B_s$ AND $\bar{B}_s$
}\bigskip
\centerline{R. Aleksan}\par
\centerline{Centre d'Etudes Nucl\'eaires de Saclay, DPhPE, 91191
Gif-sur-Yvette,
France} \par
\bigskip
 \centerline{A. Le Yaouanc, L. Oliver, O. P\`ene and J.-C. Raynal}
\par
 \centerline{Laboratoire de Physique Th\'eorique et Hautes Energies
\footnote{Laboratoire associ\'e au Centre National de la Recherche
Scientifique}}  \centerline{Universit\'e de Paris XI, b\^atiment 211, 91405
Orsay
Cedex, France}
\bigskip \bigskip
\baselineskip=14pt
\noindent
${\bf Abstract}$ \par
To help the difficult determination of the angle $\gamma$ of the unitarity
triangle,
Aleksan, Dunietz and Kayser have proposed the modes of the type
$K^-D^+_s$, common to $B_s$ and $\bar{B}_s$. We point out that it is
possible to gain in statistics by a sum over all modes with ground state mesons
in
the final state, i.e. $K^-D^+_s$, $K^{*-}D_+^s$, $K^-D^{*+}_s$,
$K^{*-}D^{*+}_s$. The delicate point is the relative phase of
these different contributions to the dilution factor $D$ of the time-dependent
asymmetry. Each contribution to $D$ is proportional to a product $F^{cb}$
$F^{ub}$ $f_{D_s}$ $f_K$ where $F$ denotes form factors and $f$ decay
constants. Within a definite phase convention, lattice calculations do
not
show any change in sign when extrapolating to light quarks the form factors and
decay
constants. Then, we can show that all modes contribute constructively to the
dilution
factor, except the $P$-wave $K^{*-}D^{*+}_s$, which is small.
Quark model arguments based on wave function overlaps also confirm this
stability in
sign. By summing over all these modes we find a gain of a factor 6 in
statistics
relatively to $K^-D^+_s$. The dilution factor for the sum
$D_{tot}$ is remarkably stable for theoretical schemes that are not in very
strong
conflict with data on $B \to \psi K(K^*)$ or extrapolated from semileptonic
charm form
factors, giving $D_{tot} \geq 0.6$, always close to $D(K^- D^+_s)$. \par

\newpage
\pagestyle{plain}
\baselineskip 18pt

Time dependent CP violating asymmetries

$$A(t) \sim D \ Im \left [ {q \over p} {\bar{M} \over M} \right ] \sin(\Delta
Mt)
\eqno(1)$$

\noindent depend on the physical quantity

$${q \over p} {\bar{M} \over M} \eqno(2)$$

\noindent which is invariant under phase redefinition of the $|B^0>$,
$|\bar{B}^0>$
states. $M$ and $\bar{M}$ are the decay amplitudes of $B^0$ and $\bar{B^0}$ to
some
common final state $|f>$~:

$$M = < f|{\cal H}_W|B^0>  \qquad 	\bar{M} = < f|{\cal H}_W|\bar{B}^0>
\eqno(3)$$

\noindent The mass eigenstates are :

$$|B_{1,2}> = p|B^0> \pm q|\bar{B}^0>				\eqno(4)$$

\noindent $D$ is the ``dilution factor" that differs from 1 for common final
states
that are not CP-eigenstates. For the moment, in the expression of the asymmetry
(1)
we have neglected the possible FSI phases, that we will discuss below. It is
obviously
important to have a large dilution factor $D$ since the number of needed pairs
$B_s$,
$\bar{B}_s$ to observe a given asymmetry scales like the $A^{-2}$ or
$D^{-2}$. \par

	We will adopt Wolfenstein phase convention and parametrization of the CKM
matrix \cite{1}. In this convention all CKM matrix elements are real except
$V_{ub}$ and $V_{td}$ and it is simple to identify which modes will contribute
to the determination of the different angles on the unitarity triangle
$\alpha$,
$\beta$ and $\gamma$. In the Standard Model we have $|q/p| = 1$ to a very good
approximation. In Wolfenstein phase convention $(q/p)_{B_d}$ is complex since
it
depends on $V_{td}$  while $(q/p)_{B_s}$ is real as it depends on $V_{ts}$. On
the
other hand, the CKM factor of the decay amplitudes is real for $b \to c$
transitions
while it is complex for $b \to u$ transitions. This gives us three different
possibilities for a non-vanishing $Im \left [{q \over p} {\bar{M} \over M}
\right]$: \par

1. $b \to u$ transitions of the $B_d$-$\bar{B}_d$ system, related to the
angle $\alpha$; \par

2. $b \to c$ transitions of
the $B_d$-$\bar{B}_d$ system, related to $\beta$, and \par

3. $b \to u$ transitions of the $B_s$-$\bar{B}_s$ system, related to $\gamma$.
\par

\noindent Of course, this is only true within Wolfenstein approximation up to
$O(\lambda^3)$,
and in the tree approximation, since also Penguin diagrams can complicate the
picture
and pollute the determination of some angles, mostly $\alpha$ and $\gamma$.
Examples
of the three types of modes, which are CP eigenstates, are respectively :
$B_d$ ,
$\bar{B}_d \to \pi^+\pi^-$~; $B_d$ , $\bar{B}_d \to J/\psi K_S$, and $B_s$,
$\bar{B}_s
\to \rho^0 K_S$ \cite{2}. \par

	There are several improvements one can think of. First, one can consider modes
that
can help to cleanly isolate the CP phase we are interested in, avoiding the
unwanted
phases coming from Penguins. Second, one can try to find \underbar{non-CP
eigenstate
common modes} to $B^0$ and $\bar{B}^0$ that, although not so clean as CP
eigenstates,
can help to increase the statistics \cite{3}. Third, one can consider
\underbar{sums
over some channels} or semi-inclusive modes that can increase the statistics if
they
contribute constructively to the asymmetry \cite{4}, \cite{5}. \par
	In this paper we will be concerned with the angle $\gamma$. Assuming unitarity
of the
CKM matrix, there are two possible determinations of $\gamma$. If $\alpha$ and
$\beta$ are measured through $B_d$, $\bar{B}_d$ decays, then $\gamma = \pi -
\alpha -
\beta$ is in principle known. However, there is an independent check, the
possible
determination outlined above through $B_s$, $\bar{B}_s$ decays, for example the
CP
eigenstate mode $\rho^0 K_S$ (Fig. 1). This measurement of $\gamma$ will be
complicated
by the expected quick $B_s$-$\bar{B}_s$ oscillations, that can wash out the CP
asymmetry. Moreover, the mode $\rho^0 K_S$ is expected to have a very small
branching
ratio ($10^{-6} - 10^{-7}$), since it is not only CKM suppressed by $V_{ub}$,
as it is
necessary, but it is also color suppressed (a further factor 0.2 in amplitude).
This
mode is also polluted by Penguin diagrams. \par

Aleksan, Dunietz and Kayser\cite{6} have proposed an alternative way of
measuring
$\gamma$, namely to consider decay modes of the type $K^-D^+_s$ that are not CP
eigenstates but are common to $B_s$ and $\bar{B}_s$, the amplitudes being
respectively
proportional to $V_{cs}V^*_{ub}$ and $V_{cb} V^*_{us}$ , both of
order $\lambda^3$ in terms of the Wolfenstein expansion parameter. This mode is
not
color suppressed, and one expects a branching ratio of $O(10^{-4})$. Moreover,
this
mode is not polluted by Penguins. Although we have here the drawback of the
dilution
factor $D$, both amplitudes $B_s$, $\bar{B}_s \to K^-D^+_s$ are of the same
order and
one can expect \cite{6} a large dilution factor for the CP asymmetry. \par

	On the other hand, we have pointed out \cite{4}, to help the determination of
the
angle $\beta$ (for which the popular CP eigenstate $\psi K_S$ is usually
proposed), to
sum over the common decay modes to $B_d$ and $\bar{B}_d$~: $D^+D^-$ ($S$-wave,
parity
violating), $D^+D^{*-} + D^{*+}D^-$ ($P$-wave, parity conserving),
$D^{*+}D^{*-}$ ($S
+ D$ waves, parity violating; $P$-wave, parity conserving). Each individual
mode,
although Cabibbo-suppressed has a decay rate of the same order as $\psi K_S$,
which is
color suppressed. We have shown, making use of the heavy-quark symmetry and
transformation properties of the weak interaction under the operator
$CPe^{i{\pi \over 2}\sigma^{(c)}_3}$ where $\sigma^{(c)}_3$ is the charm quark
spin operator
collinear to the momentum, that most of the modes (except the exchange diagram
for
$D^+D^{*-} + D^{*+}D_-$ and the $D^{*+}D^{*-}$ $P$-wave, which have small
amplitudes)
contribute constructively to the time dependent asymmetry, giving a total
dilution
factor very close to one. The gain in statistics relatively to $\psi K_S$ is of
the
order of
6, although the relative detection efficiency puts $\psi K_S$ and the sum
$D^+D^- + D^+D^{*-} + D^{*+}D^- + D^{*+}D^{*-}$ on roughly the same footing.
\par

In this paper we would like to examine the same possibility in the sum of the
modes
$K^-D^+_s$, $K^{*-}D^+_s$,$K^-D^{*+}_s$, $K^{*-}D^{*+}_s$ (and
also their CP-conjugated), that would help the difficult determination of the
angle
$\gamma$. This would be even more interesting than for the determination of
$\beta$, since we
do not have here clean modes like $\psi K_S$ of the latter case. Of course, we
do not
have here the simple situation of heavy quark symmetry. As we will see below,
assuming
factorization, each contribution to the dilution factor $D$ is proportional to
a
product $F^{cb} F^{ub} f_{D_s} f_K$ where $F$ denotes form factors
and $f$ decay constants. To the heavy-to-heavy meson form factors $F^{cb}$ and
heavy meson decay constants $f_{D_s}$ we can apply the heavy-quark symmetry
\cite{7},
but
to the heavy-to-light meson form factors $F^{ub}$ and light meson decay
constants
$f_K$ we have weaker rigorous results, like the relation between $F^{ub}$ and
$F^{uc}$ near zero recoil \cite{8}. \par

	The dilution factor $D$ is a physical quantity, independent of the phase
convention
of the states. It is convenient to work in a precise \underbar{phase
convention},
namely \underbar{the one defined by the} \par \noindent \underbar{heavy quark
symmetry} in the case of heavy quarks. For light quarks we can adopt the same
\underbar{convention} and exploit the empirical fact that the lattice
calculations,
that extrapolate from heavy masses (at the charm quark, let us say) to light
masses,
do not find changes of sign for form factors and decay constants. For example,
if
within the same phase convention $f_K$ would have a different sign than $f_D$,
then
lattice calculations would observe the quantity $f_P/f_D$ to go from 1 to zero
and
change sign when extrapolating from $P = D$ to $P = K$. This is not what is
observed
and we conclude that there is stability in sign of the form factors and decay
constants when going from heavy mesons to light mesons. This will be crucial to
have
a reliable estimation of the dilution factor $D$ when summing on the different
ground
state modes. Moreover, quark model calculations also confirm this stability in
sign.
\par

	Let us consider the final states common to $B_s$ ($\bar{b}s$) and $\bar{B}_s$
($b\bar{s}$)

\bea
|f >& =& |K^-({\vec p})D^+_s(-{\vec p}) >	\nn \\
&&|K^-({\vec p}) D^{*+}_s(\lambda = 0,- {\vec p}) > \nn \\
&&|K^{*-}(\lambda = 0,{\vec p}) D^+_s(- {\vec p}) > \nn \\
&&|K^{*-}(\lambda = 0, {\vec p})D^{*+}_s(\lambda = 0,- {\vec p}) > \nn \\
&&|K^{*-}(\lambda = \pm, {\vec p})D^{*+}_s(\lambda = \pm ,- {\vec p}) > \qquad
(5)\nn
\eea

\noindent and their CP conjugate modes :

\bea
|\bar{f} > &=& |K^+(- {\vec p})D^-_s({\vec p}) >	\nn \\
&&|K^+(- {\vec p})D^{*-}_s (\lambda = 0, {\vec p}) >	\nn \\
&&|K^{*+}(\lambda = 0,- {\vec p}) D^-_s({\vec p}) >	\nn \\
&&|K^{*+} (\lambda = 0,- {\vec p}) D^{*-}_s (\lambda = 0, {\vec p}) >	\nn \\
&&|K^{*+} (\lambda = \pm ,- {\vec p})D^{*-}_s (\lambda = \pm , {\vec p})
>\qquad					(6) \nn\eea

The spin quantization axis is along the line of flight of the decay products in
the
$B_s$ rest frame. \par

Let us write the relevant asymmetries in our case for a definite type of final
state
like $|f > = |K^-({\vec p})D^+_s(-{\vec p}) >$ and its CP conjugate mode
$|\bar{f} > = |K^+(-{\vec p}) D_s^-({\vec p}) >$. Let us call $M_{1,2}$
the amplitudes $\bar{M}(f)$, $\bar{M}(\bar{f})$ in eq. (2), respectively for
the
decays $\bar{B}_s^0 \to f$ and $\bar{B}_s^0 \to \bar{f}$ with the CKM
phases and strong phases factorized out (Fig 2) :

$$\bar{M}(f) = M_1 \qquad			\bar{M}(\bar{f}) = M_2  e^{i\gamma} e^{-i\delta_s}
\ \ \
. \eqno(7)$$

\noindent The amplitudes for the decays $B_s^0 \to \bar{f}$
and $B_s^0 \to f$  will write then

$$M(\bar{f}) =  - M_1 \qquad		M(f) = - M_2  e^{i\gamma} e^{-i\delta_s}
\eqno(8)$$

\noindent where, in Wolfenstein parametrization :

$${V^*_{ub} V_{cs} \over V^*_{us}V_{cb}}  \cong
\rho + i\eta = r e^{-i\gamma}	 \ \ \ .		\eqno(9)$$

\noindent The minus sign comes from the phase conventions
$CP|B_s^0 > = -|\bar{B}_s^0 >$, $CP|f> =|\bar{f} >$. The
phase $\gamma$ is the angle of the unitarity triangle and $\delta_s$ is a
possible strong
phase shift between both amplitudes, that we will discuss below. \par

Therefore, the quantity defined in (2) will write :

$${q \over p} {\bar{M}(f) \over M(f)}  \cong - {M_1 \over M_2} e^{i(\gamma +
\delta_s)}
\eqno(10)$$

\noindent since, within Wolfenstein phase convention, for the
$B_s^0$-$\bar{B}_s^0$ system ${q \over p}$ is approximately
real, ${q \over p} =1$, keeping only the $t$ quark in the loop. \par

The time-dependent rates are then given by \cite{2}, \cite{6}, \cite{9}:

\[R \left ( B_{phys}^0(t) \to f \right ) \sim \left ( |M_1|^2 + |M_2|^2 \right
) \left
[ 1 - R \cos (\Delta Mt) + D \sin (\gamma + \delta_s) \sin(\Delta Mt) \right ]
\]
\[R \left ( \bar{B}_{phys}^0(t) \to f \right ) \sim \left ( |M_1|^2 + |M_2|^2
\right )
\left [1 + R \cos (\Delta Mt) - D \sin (\gamma - \delta_s) \sin(\Delta Mt)
\right ] \]
\[R \left ( B_{phys}^0(t) \to \bar{f} \right ) \sim \left ( |M_1|^2 + |M_2|^2
\right )
\left [ 1 - R \cos (\Delta Mt) - D \sin (\gamma + \delta_s) \sin (\Delta Mt)
\right ] \]
\[R \left ( \bar{B}_{phys}^0 (t) \to \bar{f} \right ) \sim \left ( |M_1|^2 +
|M_2|^2
\right ) \left [ 1 + R \cos (\Delta Mt) + D \sin (\gamma - \delta_s) \sin
(\Delta Mt) \right ] (11) \]

\noindent where

$$R = {|M_1|^2 - |M_2|^2 \over |M_1|^2 + |M_2|^2} \qquad			D =
{2M_1M_2^{*} \over |M_1|^2 + |M_2|^2}	 \ \ \ . \eqno(12)$$

\noindent Note that although the weak and non-trivial strong phases have been
factorized, $M_1$ and $M_2$, following the usual conventions, can be pure
real or pure imaginary numbers according to the final state and hence the form
factors involved, as we will see below.  From expressions (11), a useful CP
asymmetry
that we can consider writes~:

$$\left \{ R \left ( B_{phys}^0(t) \to f \right ) +
R \left ( B_{phys}^0(t) \to \bar{f} \right ) \right \} -
\left \{ R \left ( \bar{B}_{phys}^0(t) \to f \right ) +
R \left ( \bar{B}_{phys}^0(t) \to \bar{f} \right ) \right \} \sim$$
$$\sim D \sin \gamma \cos \delta_s \sin (\Delta Mt)			 \eqno(13)$$

\noindent since the term in $\cos (\Delta Mt)$ cancels when summing over the
final states
$f$ and $\bar{f}$. In this expression we have two quantities affected by
hadronic
uncertainties, namely the dilution factor $D$ and the strong phase $\delta_s$.
Note that in
the asymmetry $\sin \gamma$ appears and not $\sin 2\gamma$ like it would be the
case
if both amplitudes $M_1$, $M_2$ were dependent on the same CP phase, like for
instance
the case of a final CP eigenstate. On the contrary, in our case $M_1$ and $M_2$
are
not dependent of the same CP phase, as we can see in Fig. 2. It is interesting
to note
that a curious situation could arise, namely that the asymmetry for the CP
eigenstate case $\rho^0 K_S$ (dependent on $\sin 2\gamma$) could vanish, while
the
asymmetry for the modes that we consider (dependent on $\sin \gamma$) would be
non
zero, since the region close to $\gamma = \pi/2$ is not excluded by the present
constraints on the unitarity triangle \cite{2}. \par

	We will first compute the dilution factor $D$ for each single mode and for
their sum,
and later we will discuss the possible phase $\delta_s$. \par

	Let us make a remark concerning CP violation tests using CP eigenstates or
non-CP
eigenstates. The distinction is not a deep one in the following sense. Starting
with
non-CP eigenstates like we have done above $|f>$ and its CP conjugate $|\bar{f}
> =
CP|f >$, we can always define two quantum states, eigenstates of CP :

$$|f_{\pm} > = {1 \over \sqrt{2}} |f \pm \bar{f}>			\qquad(CP = \pm )\ \ \ .
\eqno(14)$$

\noindent In this basis we can also compute the asymmetry, and we find the same
result
as above when summing over $|f_{\pm} >$ or over $|f >$, $|\bar{f} >$~:

$$\left [ R \left ( B_s^0(t) \to f_+ \right ) + R \left (
B_s^0(t) \to f^- \right ) \right ] - \left [ R \left (
\bar{B}_s^0(t) \to f_+ \right ) + R \left ( \bar{B}_s^0(t) \to f^- \right )
\right ]
=$$
$$= \left [ R \left ( B_s^0(t) \to f \right ) + R \left (
B_s^0(t) \to \bar{f} \right ) \right ] - \left [ R
\left ( \bar{B}_s^0(t) \to \bar{f} \right ) + R \left (
\bar{B}_s^0(t) \to f \right ) \right ] \eqno(15)$$

\noindent Of course, the two CP eigenstates contribute with different signs to
the
asymmetry~:

$$R \left ( B_s^0(t) \to f_+ \right ) - R \left ( \bar{B}_s^0(t) \to
f_+ \right ) =  - \left [ R \left ( B_s^0(t) \to f^- \right ) -
R \left ( \bar{B}_s^0(t) \to f^- \right ) \right ] \ \ \ . \eqno(16)$$

\noindent Working on both bases $|f_{\pm}>$ and $|f >$, $|\bar{f} >$
gives the same final result, if one sums over $|f_{\pm} >$  or
$|f >$, $|\bar{f} >$. However, unlike $|f >$, $|\bar{f} >$ the CP
eigenstates $|f_{\pm}>$ are not asymptotic states. \par

	Let us now compute the amplitudes $M_1$ and $M_2$.  For the spectator
contributions,
these amplitudes correspond respectively to the emission of a $K$ or a $D_s$
(Figs. 2a
and 2b). There are also contributions from the exchange diagrams (Figs. 3a and
3b).
The latter are quite small, following the same arguments exposed in ref. 4 (a
color
factor suppression of 0.2 in amplitude, and a form factor suppression), and we
will
neglect them. \par

	From the definitions

\[ \left < P(p)|A_\mu|0 \right > = - i f_P \  p_\mu \]
\[ \left < V(p, \lambda )|V_\mu|0 \right > = m_V f_V \epsilon^*_\mu(\lambda )
\]
\[ \left < P_i|V_\mu|P_j \right > = \left ( p^i_\mu + p^j_\mu - {m^2_j - m^2_i
\over q^2}
q_\mu \right ) f_+(q^2) + {m^2_j - m^2_i \over q^2}  q_\mu f_0 (q^2) \]
\[ \left < V_i|A_\mu|P_j \right > = \left ( m_i  + m_j \right )
A_1 (q^2) \left ( \epsilon^*_\mu - {\epsilon^*.q \over q^2} q_\mu \right ) -
A_2(q^2) {\epsilon^*.q \over
m_i + m_j} \left ( p^i_\mu + p^j_\mu - {m^2_j - m^2_i \over q^2}  q_\mu \right
) \]\[+ 2
m_i A_0(q^2)  {\epsilon^*.q \over q^2}  q_\mu \]
\[ \left < V_i|V_\mu|P_j \right > = {2
V(q^2) \over m_i + m_j} i \ \epsilon_{\mu\nu\rho\sigma} p^\nu_j
p^\rho_i \epsilon^{*\sigma}	(17)
$$

\noindent we find

$$M_1 \left ( \bar{B}_s^0 \to K^-({\vec p})D_s^+(- {\vec p}) \right ) = -
|V^*_{us} V_{cb}|  {G \over \sqrt{2}}  i \ f_K \left ( m^2_B - m^2_D \right )
f_0^{cb} (m^2_K) a_1$$

$$M_2 \left ( \bar{B}_s^0 \to K^+(-{\vec p}) D_s^0({\vec p}) \right ) = -
|V^*_{cs}
V_{ub}| {G \over \sqrt{2}} i \ f_D \left ( m^2_B - m^2_K \right )
f^{ub}_0(m^2_D)
a_1$$

$$M_1 \left ( \bar{B}_s^0 \to K^{*-}(\lambda = 0, {\vec p}) D^+_s(- {\vec p})
\right )
= + |V^*_{us} V_{cb}|  {G \over \sqrt{2}}  2 f_{K^*} m_B f_+^{cb}(m^2_{K^*})
a_1 p$$

$$M_2 \left ( \bar{B}_s^0 \to K^{*+} (\lambda = 0,-{\vec p}) D^-_s({\vec p})
\right ) =
- |V^*_{cs}V_{ub}|  {G \over \sqrt{2}}  2 f_D m_BA_0^{ub} (m^2_D) a_1 p$$

$$M_1 \left ( \bar{B}_s^0 \to K^-({\vec p}) D^{*+}_s(\lambda = 0,-{\vec p})
\right ) = -
|V^*_{us} V_{cb}| {G \over \sqrt{2}}  2 f_K m_B A_0^{cb} (m^2_K) a_1 p$$

$$M_2 \left ( \bar{B}_s^0 \to K^+(-{\vec p}) D^{*-}_s(\lambda = 0,{\vec p})
\right ) = +
|V^*_{cs} V_{ub}|  {G \over \sqrt{2}}  2 f_{D^*} m_B f_+^{ub} (m^2_{D^*}) a_1
p$$

\[M_1 \left ( \bar{B}_s^0 \to
K^{*-}(\lambda = 0,{\vec p}) D^{*+}_s(\lambda = 0,-{\vec p}) \right ) =
|V^*_{us} V_{cb}|  {G \over\sqrt{2}}  m_{K^*} f_{K^*} \]\[\left [ (m_B +
m_{D^*}) \left (
{p^2 + E_{D^*} E_{K^*} \over m_{D^*} m_{K^*}} \right )  A_1^{cb}(m^2_{K^*}) -
{m^2_B \over
m_B + m_{D^*}}  {2p^2 \over m_{D^*} m_{K^*}} A_2^{cb}(m^2_{K^*}) \right ] a_1\]

\[M_2 \left ( \bar{B}_s^0 \to
K^{*+} (\lambda = 0,-{\vec p}) D^{*-}_s(\lambda = 0,{\vec p}) \right ) =
|V^*_{cs} V_{ub}|  {G \over \sqrt{2}}  m_{D^*} f_{D^*} \]\[
	\left [ (m_B + m_{K^*}) \left (
{p^2 + E_{D^*}E_{K^*} \over m_{D^*} m_{K^*}} \right )
A_1^{ub}(m^2_{D^*}) - {m^2_B \over m_B + m_{K^*}} {2p^2 \over m_{D^*} m_{K^*}}
A_2^{ub} (m^2_{D^*}) \right ] a_1\]

\[M_1^{pv} \left ( \bar{B}_s^0 \to K^{*-}(\lambda = \pm 1, {\vec p})
D^{*+}_s(\lambda = \pm 1,-{\vec p}) \right ) =
|V^*_{us} V_{cb}|  {G \over \sqrt{2}}  m_{K^*} f_{K^*}\]\[
			(m_B + m_{K^*}) A_1^{cb} (m^2_{K^*}) a_1\]

\[M_2^{pv} \left ( \bar{B}_s^0 \to K^{*+} (\lambda = \pm1,-{\vec p}) \right )
D_s^{*-}
(\lambda = \pm 1,{\vec p}) = |V^{*}_{cs} V_{ub}|  {G \over \sqrt{2}}  m_{D^*}
f_{D^*}\]\[
		(m_B + m_{D^*}) A_1^{ub} (m^2_{D^*}) a_1\]

\[	M_1^{pc} \left ( \bar{B}_s^0 \to K^{*-}(\lambda = \pm 1, {\vec
p})D^{*+}_s(\lambda
= \pm 1,- {\vec p}) \right ) = \pm |V^*_{us} V_{cb}|  {G \over \sqrt{2}}
m_{K^*}f_{K^*}\]\[
{m_B \over m_B + m_{D^*}} 2 V^{cb}(m^2_{K^*}) a_1 p\]

\[M_2^{pc} \left ( \bar{B}_s^0 \to K^{*+}(\lambda = \pm 1,- {\vec p})
D^{*-}_s(\lambda
= \pm 1, {\vec p}) \right ) = \pm |V^*_{cs} V_{ub}| {G \over \sqrt{2}}  m_{D^*}
f_{D^*}\]
$${m_B \over m_{B^+} m_{K^*}} 2 V^{ub} (m^2_{D^*}) a_1 p				\eqno(18)$$

\noindent In these relations $p = |{\vec p}|$ is the modulus of the momentum in
the
center-of-mass, and the weak and strong phases defined in (7)-(8) have been
factorized out, except for the trivial but crucial phases that arise from the
form
factors involved. \par

We use the phase \underbar{convention} of the heavy quark symmetry. In this
phase
convention, one has the well-known heavy quark symmetry relations~:

$$f_{D^*} = f_D							\eqno(19)$$

$${2 \sqrt{m_B m_D} \over \left ( m_B + m_D \right )} f^+(q^2) = {2 \sqrt{m_B
m_D}
\over \left ( m_B + m_D \right )} {f_0(q^2) \over \left [ 1 - {q^2 \over \left
( m_B
+ m_D \right )^2} \right ]}  = {2 \sqrt{m_B m_{D^*}} \over \left ( m_B +
m_{D^*}
\right )} V(q^2) =$$	 $$= {2 \sqrt{m_B m_{D^*}} \over \left ( m_B + m_{D^*}
\right )}
A_0(q^2) = {2 \sqrt{m_B m_{D^*}} \over \left ( m_B + m_{D^*} \right )} A_2(q^2)
=
{2 \sqrt{m_B m_{D^*}} \over \left ( m_B + m_{D^*} \right )} {A_1(q^2) \over
\left [ 1 -
{q^2 \over \left ( m_B + m_{D^*} \right )^2} \right ]}  = \xi(v.v') \eqno(20)$$

\noindent that, in our problem, apply to the $b \to c$ form factors and, within
the
factorization approximation, to the emission of $D^+_s$, $D^{*+}_s$. \par

For light mesons we will adopt the same phase convention and assume, as
suggested by
lattice calculations, that the sign of the different form factors and decay
constants
does not change when extrapolating from heavy lo light mesons. This means that
we
assume the same relative signs as implied by the preceding equations above,
e.g.
$f_{K^*}/f_K > 0$, $A_1(q^2)/f_0(q^2) > 0$, etc., but we do not necessarily
assume the heavy quark symmetry ratios (although we could make such hypothesis
as a
possible Ansatz, see below). \par

If we consider all the modes $PP$, $PV$ and $VV$, and we neglect the strong
phase of
each mode, the asymmetry has the same form as in (13), with the dilution factor
$D$
given by

$$D = {2\sum_i M^{(i)}_1 M^{(i)*}_2 p(i) \over \sum_i
\left [ |M^{(i)}_1|^2 + |M^{(i)}_2|^2 \right ] p(i)}
\eqno(21)$$

\noindent where the sum extends over all the modes enumerated above with
momenta
$p(i)$. \par

To be clear, we will give the different contributions to the numerator, $M_1(
\bar{B}_s^0 \to f)$ $M_2^*(\bar{B}_s^0 \to \bar{f})$ with their crucial sign,
the
denominator being obvious.

$$|f> = |K^-({\vec p})D^+_s(- {\vec p})>		\quad |\bar{f}> =
|K^+(- {\vec p}) D^-_s(p)>$$
$$M_1 M_2^* = \left | V^*_{ub} V_{cs} V^*_{us} V_{cb} \right | {G^2 \over
\sqrt{2}}
f_K \left ( m^2_B - m^2_D \right )  f^{cb}_0 (m^2_K) f_D \left ( m^2_B - m^2_K
\right
) f^{ub}_0 (m^2_D) (a_1)^2$$

$$|f > = |K^{*-}(\lambda = 0, {\vec p}) D^+_s(- {\vec p}) > \quad	|\bar{f}> =
-|K^{*+}(\lambda = 0,- {\vec p}) D^-_s({\vec p})>$$
$$M_1 M_2^* = \left | V^*_{ub} V_{cs} V^*_{us} V_{cb} \right | {G^2 \over
\sqrt{2}} 4
f_{K^*} m_B f_+^{cb} (m^2_{K^*}) f_D m_B A^{ub}_0 (m^2_D) p^2 (a_1)^2$$

$$|f > = |K^-({\vec p}D^{*+}_s(\lambda = 0,-{\vec p}) >	\quad |\bar{f}> =
-|K^+(- {\vec p} D^{*-}_s(\lambda = 0, {\vec p})>$$
$$M_1 M_2^* = \left | V^*_{ub} V_{cs} V^*_{us} V_{cb} \right |{G^2 \over
\sqrt{2}} 4
f_K m_B A^{cb}_0 (m^2_K) f_{D^*} m_B f_+^{ub}(m^2_{D^*}) p^2 (a_1)^2$$

$$|f > = |K^{*-}(\lambda = 0, {\vec p})D^{*+}_s (\lambda = 0,- {\vec p})>	\quad
|\bar{f}> =|K^{*+}(\lambda = 0,- {\vec p})D^{*-}_s (\lambda = 0, {\vec p})>$$
$$M_1 M_2^* = \left | V^*_{ub} V_{cs} V^*_{us} V_{cb} \right | {G^2 \over
\sqrt{2}}
m_{K^*} f_{K^*} m_{D^*} f_{D^*}$$
$$\left [ \left ( m_B + m_{D^*} \right ) \left ( {p^2 + E_{D^*} E_{K^*} \over
m_{D^*}
m_{K^*}} \right )  A_1^{cb} (m^2_{K^*}) - {m^2_B \over m_B + m_{D^*}}  {2p^2
\over m_{D^*}
m_{K^*}} A_2^{cb} (m^2_{K^*}) \right ]$$
$$\left [ \left (m_B +m_{K^*} \right ) \left ( {p^2 + E_{D^*} E_{K^*} \over
m_{D^*}
m_{K^*}} \right ) A_1^{ub} (m^2_{D^*}) - {m^2_B \over m_B  + m_{K^*}}
{2p^2 \over m_{D^*} m_{K^*}} A_2^{ub} (m^2_{D^*}) \right ] (a_1)^2$$

$$|f > = |K^{*-}(\lambda = \pm 1, {\vec p}) D^{*+}_s(\lambda = \pm 1,-{\vec
p})> \quad
|\bar{f}> = |K^{*+}(\lambda = \pm 1,-{\vec p}) D^{*-}_s (\lambda = \pm 1, {\vec
p})>$$
\[M_1^{pv} M_2^{pv*}= \left | V^*_{ub} V_{cs} V^*_{us} V_{cb} \right | {G^2
\over
\sqrt{2}} m_{K^*} f_{K^*} m_{D^*} f_{D^*}	\left ( m_B + m_{K^*} \right )
A_1^{cb}
(m^2_{K^*}) \]\[\left ( m_B + m_{D^*} \right )  A_1^{ub} (m^2_{D^*}) (a_1)^2\]

$$|f > = |K^{*-}(\lambda = \pm 1, {\vec p}) D^{*+}_s)(\lambda =  \pm 1,- {\vec
p})>
\quad |\bar{f} > = |K^{*+}(\lambda = \pm 1,- {\vec p}) D^{*-}_s)(\lambda = \pm
1, {\vec
p})>$$
\[M_1^{pc} M_2^{pc*} = - \left | V^*_{ub} V_{cs} V^{*}_{us} V_{cb} \right |
{G^2 \over
2} m_{K^*} f_{K^*} m_{D^*} f_{D^*} 4 {m_B \over m_B + m_{D^*}}\]
$$ V^{cb}(m^2_{K^*}) {m_B \over
m_B + m_{K^*}} V^{ub}(m^2_{D^*})  p^2 (a_1)^2			\ \ \ . \eqno(22)$$

\noindent Taking the sign of the $KD_s$ contribution as reference, we see that
only the term with $K^*D^*_s$ in the $P$-wave (parity conserving) contributes
with a negative sign. However, we need also to discuss the longitudinal (parity
violating) $K^*D^*_s$ because there can be cancellations in its expression.
On the other hand, following our conventions for the $K^\ast D$ and $K D^\ast$
cases, note the minus
sign of
the CP conjugate states, that combines with the opposite sign of this case in
(18) to
give a constructive sign also for these modes. Finally, let us remark that in
the
limit in which we consider the $s$ quark as heavy, we recover the relative
signs of
the spectator diagram of the $B_d \to D\bar{D}$, ... case studied in ref. 4, as
it
should. \par

	Owing to the heavy flavor symmetry relations, the heavy quark transition
parenthesis
is positive~:

$$\left [ \left ( m_B + m_{D^*} \right ) \left (
{p^2 + E_{D^*} E_{K^*} \over m_{D^*} m_{K^*}} \right )
A_1^{cb} (m^2_{K^*}) -
{m^2_B \over m_B + m_{D^*}}
{2p^2 \over m_{D^*} m_{K^*}}
A_2^{cb} (m^2_{K^*}) \right ] > 0  \ \ \ . \eqno(23)$$

\noindent However, we do not have any sound theoretical basis to claim a
definite sign
for the heavy-to-light parenthesis~:

$$\left [ \left ( m_B + m_{K^*} \right ) \left ( {p^2 + E_{D^*} E_{K^*} \over
m_{D^*}
m_{K^*}} \right ) A_1^{ub} (m^2_{D^*}) - {m^2_B \over m_B + m_{K^*}} {2p^2
\over m_{D^*}
m_{K^*}} A_2^{ub} (m^2_{D^*}) \right ]  \ \ \ . 	\eqno(24)$$

	The precise value of the dilution factor (21) is rather uncertain with the
present
knowledge of heavy-to-light form factors. One must wait for precise data on
semileptonic decays $B \to \pi(\rho) \ell \nu$  to estimate $D$, assuming
factorization. \par

	However, one can try to constrain the form factors $f_+^{ub}$,
$V^{ub}$, $A_1^{ub}$, $A_2^{ub}$ from data concerning other heavy-to-light
quark
transitions. First, we have data on non-leptonic decays like the $B \to \psi
K$, $\psi
K^*$ transitions.  These decays depend on $b \to s$ form factors that, assuming
factorization, could be related to our case by light flavor $SU(3)$ symmetry.
Second,
we have also data on the semileptonic form factors $D \to K(K^*)\ell\nu$, and
we can
try to extrapolate to the $b \to s$ form factors using then the heavy flavor
symmetry \cite{8}. \par
	Recent data on $B \to \psi K$, $\psi K^*$ (the rates and the ratio
$\Gamma_L/\Gamma_{tot}$ for
the latter) are hardly compatible with current models of non-leptonic $B$ meson
decays
and with the present knowledge of heavy-to-light form factors. We have
performed this
analysis in detail in \cite{10} (see also Gourdin, Kamal and Pham \cite{11}).
We will make
here a brief resum\'e of our discussion \cite{10} to be able to estimate the
dilution factor
$D$ (21), the main object of the present paper. Let us consider the ratio of
rates

$$R = {\Gamma \left ( \bar{B}_d^0 \to
\psi K^{*0} \right ) \over \Gamma \left ( \bar{B}_d^0 \to
\psi K^{0} \right )}				 \eqno(25)$$

\noindent and the polarization ratio for $\psi K^{*0}$ (L stands for
longitudinal
polarization)~:

$$R_L = {\Gamma_L \left ( \bar{B}_d^0 \to
\psi K^{*0} \right ) \over \Gamma_{tot} \left ( \bar{B}_d^0 \to \psi K^{*0}
\right )}
\eqno(26)$$

%___________________________________________________________________________
\begin{table}
\centering
\begin{tabular} {|c|c|c|}
\hline
&{$R$} &{$R_L$} \\ \hline
{Exp.} & \ $1.64 \pm 0.34$ &$0.78 \ (ARGUS)\cite{17}$  \\
 & &$0.80 \pm 0.08 \pm 0.05 \ (CLEO)\cite{17}$ \\
 & &$0.66 \pm 0.10\matrix{& + 0.10 \cr &-0.08 \cr} \ (CDF)\cite{18}$
\\  \hline
BSWI$\cite{12}$ &{4.23} &0.57 \\  \hline
BSWII$\cite{14}$ &{1.61} &0.36 \\  \hline
GISW$\cite{15}$ &{1.71} &0.06 \\  \hline
QCDSR$\cite{16}$ &{7.60} &0.36 \\  \hline
\end{tabular}
\caption{\it The ratios $R$ and $R_L$ defined in the text compared to models of
$B \to K(K^*)$ form factors.}
\end{table}
%---------------------------------------------------------------------------

The data on $R$ and $R_L$ are given in Table 1, together with the predictions
of the
models of Bauer, Stech and Wirbel on non-leptonic $B$ decays. BSWI stands for
the
version of the model in which all form factors are pole-like \cite{12},
\cite{13}, and BSWII
for the latter version \cite{14} where the form factor $A_1(q^2)$ has a pole
shape, and
the rest of the form factors, namely $f_+(q^2)$, $V(q^2)$, $A_2(q^2)$, have a
dipole
one, the ratio between both types being a pole, like in the heavy-to-heavy
case. We
consider only the ratios of rates because the absolute magnitude depends on the
effective color factor $a_2$ (the modes under consideration are of the class
II, color
suppressed), which is fitted from the data \cite{12}, \cite{14}. We present the
different
values of $R_L$ from the experiments ARGUS, CLEO and CDF. We see that both
descriptions of $R$ and $R_L$ are not satisfactory, either $R$ is too large or
$R_L$
too small by roughly a factor 2. For completeness, we give the predictions of
the form
factors of the GISW model \cite{15}, that gives a very small value for $R_L$,
and of
QCD sum rules \cite{16}, provided to us by P. Ball. \par

	We have also examined \cite{10} the possibility of the extrapolation from the
data on
$D \to K(K^*)$ form factors to the $B \to K(K^*)$ form factors using heavy
quark
symmetry scaling laws \cite{8} in the region of $q^2$ near
$q^2_{max}$. Notice that the non-leptonic data give us information at
a different kinematic point (at $q^2 = m^2_{\psi})$ than the data on
semileptonic $D$ decays (at small $q^2$) or the heavy quark limit QCD scaling
laws (at
$q^2_{max}$). Therefore, going from semileptonic $D$ decays to
non-leptonic $B \to \psi K(K^*)$ decays requires a double extrapolation, from
$q^2 =
0$ to $q^2_{max}$ in $D \to K(K^*)$, and from $q^2_{max}$ in
$B \to K(K^*)$ to $q^2 = m^2_{\psi}$ in $B$ non-leptonic decays. These
extrapolations are obviously very sensitive to the proposed Ansatz for the
$q^2$
dependence of the form factors, and to the corrections to the heavy quark limit
scaling laws near $q^2_{max}$. \par
	For the form factors $D \to K(K^*)$ at $q^2 = 0$ we have the world
average \cite{19}~:

\[f^{sc}_+(0) = 0.77 \pm 0.04 \]
\[V^{sc}(0) = 1.16 \pm 0.16	\]
\[A^{sc}_1(0) = 0.61 \pm 0.05	\]
\[A^{sc}_2(0) = 0.45 \pm 0.09	(27)
$$

\noindent The ratios $R$ and $R_L$ depend on the ratios of $B \to K(K^*)$ form
factors
$f^{sb}_+(m^2_{\psi})/A^{sb}_1(m^2_{\psi})$,
$V^{sb}(m^2_{\psi})/A^{sb}_1(m^2_{\psi})$,
$A^{sb}_2(m^2_{\psi})/A^{sb}_1(m^2_{\psi})$. If we impose the scaling laws $B
\to D$
at $q^2_{max}$, plus an Ansatz for the $q^2$ dependence of the ratios of
form factors $f_+(q^2)/A_1(q^2)$, $V(q^2)/A_1(q^2)$,
$A_2(q^2)/A_1(q^2)$, we can also consider $R$ and $R_L$ as functions of the
ratios $f^{sc}_+(0)/A^{sc}_1(0)$,
$V^{sc}(0)/A^{sc}_1(0)$,
$A^{sc}_2(0)/A^{sc}_1(0)$. Then we can make a $q^2$ fit
to the data on $R$, $R_L$ and the ratios of $D \to K(K^*)$ form factors at $q^2
= 0$.
As discussed at length in \cite{10}, we can assume several hypothesis
concerning the
relevant behaviors: \par
1) Scaling laws at $q^2_{max}$~; \par
2) $q^2$ behavior of the ratios $f_+(q^2)/A_1(q^2)$, $V(q^2)/A_1)(q^2)$,
$A_2(q^2)/A_1)(q^2)$~; \par
3) absolute value and $q^2$ dependence of the form
factors, let us say $A_1(q^2)$. \par
The cases that we will consider here are~: \par
{\bf 1. Scaling laws at $q^2_{max}$.} \par We consider two cases. \par
\hskip 1 truecm {\bf 1.1. Asymptotic
scaling laws.} There is a constraint from the heavy quark symmetry that relates
the form factors, say $D \to K$ and $B \to K$ near zero recoil ${\vec q} = 0$
(i.e. at
$q^2_{max}$ for each process) \cite{8}~:

$${f^{sb}_+(q^2_{max}) \over f^{sc}_+(q^2_{max})} = {V^{sb}(q_2^{max}) \over
V^{sc}(q^2_{max})} = {A^{sb}_2(q^2_{max}) \over A^{sc}_2(q^2_{max})} = \left (
{m_B
\over m_D} \right )^{{1 \over 2}}$$
$${A^{
sb}_1(q^2_{max} \over A^{sc}_1(q^2_{max})}
= \left ( {m_D \over m_B} \right )^{{1 \over 2}} \ \ \ . \eqno(28)$$

\hskip 1 truecm {\bf 1.2. Softened scaling, i.e. scaling law smoothed by mass
corrections.}
	Motivated by the pole factor in the Isgur-Wise relations (20), that will be
shown to
appear also in the heavy-light case in a quark model of meson form factors
\cite{20},
we will assume also another type of extrapolation, namely a constant form
factor
$A_1(q^2)$ and a single pole for $f_+(q^2)$ (at $q^2 = (m_B + m_K)^2$ or $(m_D
+
m_K)^2$ according to the transition), and also for $V(q^2)$, $A_2(q^2)$ (at
$q^2 =
(m_B + m_{K^*})^2$ or $(m_D + m_{K^*})^2$). Interestingly, if one does not
neglect the
light meson masses, this type of extrapolation leads to a smoothed scaling law
that
exhibits rather large corrections to the asymptotic scaling (28)~:

$${f^{sb}_+(q^2_{max}) \over f^{sc}_+(q^2_{max})} = \left ( {m_D \over m_B}
\right
)^{{1 \over 2}} \left ( {m_B + m_K \over m_D + m_K} \right )$$

$${V^{sb}(q^2_{max}) \over V^{sc}(q^2_{max})} = {A^{sb}_2(q^2_{max}) \over
A^{sc}_2(q^2_{max})} = \left ( {m_D \over m_B} \right )^{{1 \over 2}}
\left ({m_B + m_{K^*} \over m_D + m_{K^*}} \right)$$

$${A^{sb}_1(q^2_{max}) \over A^{sc}_1(q^2_{max})} = \left ( {m_B \over m_D}
\right
)^{{1 \over 2}} \left ( {m_D + m_{K^*} \over m_B + m_{K^*}} \right )
\eqno(29)$$

{\bf 2. $q^2$ dependence of the ratios $f_+(q^2)/A_1(q^2)$,
$V(q^2)/A_1(q^2)$, $A_2(q^2)/A_1(q^2)$.} \par
Concerning this second question, several models
have been proposed: a single pole for all form factors \cite{12} or a single
pole for
$A_1(q^2)$ and a dipole for $f_+(q^2)$, $V(q^2)$, ${A_2(q^2)}$ \cite{14}. The
latter
model is inferred from applying the heavy-to-heavy symmetry relations (19) to
the
heavy-to-light case and assuming a pole form factor for $A_1(q^2)$. Both
possibilities
correspond to the cases: \par {\bf 2.1. Constant behaviour of the ratio.}  \par
{\bf 2.2. Pole
behavior of the ratio.} \par

{\bf 3. $q^2$ dependence of $A_1(q^2)$. }\par
Here we will consider two possibilities : {\bf 3.1.
Pole behavior.} \par {\bf 3.2. Constant.} This is a simplifying situation
suggested by
our quark model \cite{20} that predicts a weak $q^2$ dependence for this form
factor.
\par

	In \cite{10} we argue that the case of $f_+$ is special in the sense that one
has a
quasi-Goldstone boson in the final state and we begin first with a fit to the
ratio
$R_L$ and the ratios of form factors $V^{sc}(0)/A^{sc}_1(0)$,
$A^{sc}_2(0)/A^{sc}_1(0)$, that concern $D \to K^*$ alone,
assuming the several extrapolation hypothesis, corresponding to the
combinations of
scaling and $q^2$ dependence 1.1, 1.2, 2.1, 2.2 described above, corresponding
to
asymptotic or softened scaling and to a constant or pole ratio of form factors.
In
this case we have 3 data and 2 parameters (the ratios of form factors) and
hence one degree of freedom:
${DoF} = 1$. The fit is very bad for the CLEO and ARGUS data $(\chi^2/DoF >
9$), but
for the CDF data the cases of extrapolation 1.2 (softened scaling) together
with 2.1
(constant ratio) or with 2.2 (pole ratios) are qualitatively favored (see Table
2,
where we only report the fit to CDF data), especially the former. The best fit
corresponds to softened scaling with pole ratios as a function of $q^2$. \par

%___________________________________________________________________________
\begin{table}
\centering
\begin{tabular} {|c|c|c|c|c|}
\hline
  &   	  	       	 		        $R_L$ &            $\frac{V^{sc}(0)}
{A_1^{sc}(0)}$ &  $\frac{A_2^{sc}(0)} {A_1^{sc}(0)}$  &	  $\chi^2/DoF$ \\
\hline
	Exp.   &   $0.66 \pm 0.14 $   &$  1.90 \pm 0.25 $&$     0.74 \pm 0.15$&\\
\hline
	\multicolumn{5}{|c|}
	{Extrapolation $D\to B$} \\
	(I)	 &       0.23	&   1.66        &        0.37	     &            16.1
\\	(II)	 &       0.32	&   1.70        &        0.48	     &              9.5
\\	(III)     &       0.39	&   1.75         &       0.56	     &              5.7
\\	(IV)	   &     0.49	&   1.82        &        0.66	     &              1.9
\\  \hline
\end{tabular}
\caption{\it Fit to the ratio $R_L$ and to the ratios of form factors $D
\to K^*$. Extrapolation corresponds to asymptotic scaling with constant (I), or
pole
(II) ratios, or to softened scaling with constant (III) or pole (IV) ratios
(see the
text). For $R_L$ only CDF data are included. For ARGUS [17] ($R_L > 0.78$) or
CLEO [17]
($R_L = 0.80 \pm 0.08 \pm 0.05$) data, one has
$\chi^2/DoF > 9$ in all cases. }
\end{table}
%---------------------------------------------------------------------------

	We then make in ref. \cite{10} an overall $q^2$ fit to the ratios $R$, $R_L$
and the
ratios
of form factors $f_+^{sc} (0)/A^{sc}_1(0)$,
$V^{sc}(0)/A^{sc}_1(0)$,
$A^{sc}_2(0)/A^{sc}_1(0)$ (we have now ${DoF} = 2$)
assuming several extrapolation hypothesis. The $\chi^2/DoF$ turns out to be
very
bad in
all these cases if we stick to the CLEO or ARGUS data for $R_L$ ($\chi^2/DoF
> 8$). We
show in Table 3 only the results of the fit in the case of CDF, although in our
detailed discussion \cite{10} the CLEO data are included. It is clear from
Table 3 that
the fits are not good, but we give the whole set as it is useful to see how the
different hypothesis compare to the data. Reasonable fits with comparable
values of
$\chi^2$ are obtained with softened scaling and constant (III) or pole ratio of
form
factors (IV), and also asymptotic scaling with pole ratio (II), if we accept
the CDF
data. We compute the dilution factors for all these cases. \par

%___________________________________________________________________________
\begin{table}
\centering
\begin{tabular} {|c|c|c|c|c|c|c|}
\hline
  &   	  	     $R$&  	 		        $R_L$ &  $\frac{f_+^{sc}(0)} {A_1^{sc}(0)}$ &
        $\frac{V^{sc}(0)} {A_1^{sc}(0)}$ &  $\frac{A_2^{sc}(0)} {A_1^{sc}(0)}$
&	  $\chi^2/DoF$ \\  \hline
	Exp.   & $1.64\pm0.14$&  $0.66 \pm 0.14 $ &$1.26\pm0.12$  &$  1.90 \pm 0.25
$&$     0.74 \pm 0.15$&\\  \hline \multicolumn{7}{|c|}
	{Extrapolation $D\to B$}  \\
	(I)&1.44 &       0.23	&  1.20     &        1.79     &           0.34	&
8.5
\\	(II)&1.60	 &       0.32	&      1.25    &         1.72    &            0.47	&
      4.7
\\	(III)&1.81     &       0.37 &	   1.32     &        1.66       &         0.60
 &      3.2
\\	(IV)&2.15	   &     0.45	&   		   1.45    &         1.62    &            0.81
 &      4.2
\\  \hline
\end{tabular}
\caption{\it  Fit to the ratios $R$, $R_L$ and to the ratios of form
factors $D \to K^*$. Extrapolation corresponds to asymptotic scaling with
constant
(I), or pole (II) ratio, or to softened scaling with constant (III) or pole
(IV)
ratio (see the text). For $R_L$ only CDF data are included. For ARGUS [17]
($R_L >
0.78$) or CLEO [17] ($R_L = 0.80 \pm 0.08 \pm 0.05$) data, one has $\chi^2/DoF
> 8.5$ in
all cases.   }
\end{table}
%---------------------------------------------------------------------------

	Concerning the CP asymmetries considered in this paper, and in order to be
free of a
too strict model dependence, we will compute $D$ in the various theoretical
schemes,
assuming $SU(3)$ symmetry. In Table 4 we present the predictions for the
dilution
factor  $D_{K^- D^+_s}$ of the mode
$K^-D^+_s$, the total dilution factor $D_{tot}$, given by
(21), and the statistical gain $\Gamma_{tot}/\Gamma(K^-D^+_s)$ if
we consider the sum over the ground state modes in the asymmetry. For the above
discussed extrapolation schemes we retain only those that are not in a too
strong
conflict with the data on $B \to \psi K(K^*)$, as explained in Table 4. For all
relatively reasonable schemes, the sign of the parenthesis (24) is positive,
like for
the heavy-to-heavy case. The parameters that we have adopted are the recent
values
$|V_{cb}|=0.037$, $|V_{ub}|=0.003$ (corresponding to $|V_{ub}|/|V_{cb}|=0.08$),
$\tau_B=1.49\,10^{-12}$ sec, and for the effective QCD factor of class I
decays, $a_1 = 1.15$. The Isgur-Wise function at the $K$ or $K^\ast$ mass
corresponds roughly to $\xi(1.5) \simeq
0.60$. We take the decays constants $f_{D_s} = 0.23$ GeV and
$f_{D_s^\ast} = 0.28$ GeV, as suggested by lattice calculations \cite{21}.
	Although there is not any satisfying scheme, we conclude from our calculations
that
theoretical schemes that do not give $R_L$ too small or/and $R$ too large also
give a
dilution factor $D_{tot} \geq 0.6$, and that we obtain a dilution factor for
the sum of all
final states that is close to the case $K^-D_s^+$, and with a
statistical gain of the order of 5-6. Notice that we find somewhat smaller
values for
the dilution factor $D(K^-D_s^+)$ than in \cite{6}. This can be
traced back to the present smaller value of $|V_{ub}|/|V_{cb}|$ and the more
precise value of
the Isgur-Wise function.

%___________________________________________________________________________
\begin{table}
\centering
\begin{tabular} {|c|c|c|c|c|}
\hline
  &   	  	     $D_{K^-D_s^+}$&  	 		        $D_{tot}$ &  $BR(K^-D_s^+)$ &
   $\Gamma_{tot}/\Gamma(K^-D_s^+)$ \\  \hline
	BSWI	       &       0.61     &        0.57	 &   $3.5 \times 10^{-4 }$&
5.5\\
BSWII     &         0.61   &          0.58	&   $3.5 \times 10^{-4 }$&
   5.5
\\	GISW	&	    0.06    &         0.02     &        $1.1 \times 10^{-4 }$&	5.8
\\	QCDSR	&    0.42	&    0.52      &     $6.0 \times 10^{-4 }$&	6.0
\\	Lattice (a) &    $0.67 \pm 0.12$&$	 0.60 \pm 0.15	$&$   (3.6 \pm 1.9)\times
10^{-4}$&$      8.9 \pm 1.3$
\\  	Lattice (b)  &   $0.67 \pm 0.12$&$	 0.60 \pm 0.15	$&$   (3.6 \pm
1.9)\times 10^{-4}$&$      8.8 \pm 1.4$
\\   \hline
Extrapolation $B\to D$ &&&& \\
(III)	   & 	    0.51 	&     0.56	&    $3.3 \times 10^{-4 }$	   &      5.8 \\
(IV.1)	&    0.55 &	     0.51&	    $3.4 \times 10^{-4 }$	&     5.5\\
(IV.2)	&    0.84 &	     0.74&$4.0 \times 10^{-4 }$   &      5.3\\   \hline
\end{tabular}
\caption{\it  Dilution factor for the mode
$K^-D^+_s$ and for the sum of all ground state mesons $D_{tot}$
and total statistical gain for different theoretical schemes of Tables 1 and 2.
Lattice results correspond to a pole shape for $A_1(q^2)$ and a constant (a) or
a pole
(b) ratio of form factors. The extrapolations correspond to the values of form
factors of the best fits in Table 3 plus the central value $A_1(0) = 0.61$ from
experiment : softened scaling law plus constant ratio between the form factors
with
pole shape for $A_1(q^2)$ (III) or softened scaling law plus pole ratio between
the
form factors with pole shape for $A_1(q^2)$ (IV.1) or taking $A_1(q^2)$ to be a
constant (IV.2). The case (IV.2) is the favored one on phenomenological grounds
(see
the text).   }
\end{table}
%---------------------------------------------------------------------------

	A last few comments on the results of Table 4. We should emphasize that the
data for
$f_+(q^2)$ seem to favor a single pole \cite{19} at the mass of the
corresponding $1^-$ state and
therefore the cases (III) or (IV.2) corresponding to constant or pole ratio of
form
factors with $A_1(q^2)$ a pole or a constant are favored from this point of
view, while
case (IV.1), which assumes a dipole shape for $f_+(q^2)$ and a pole for
$A_1(q^2)$ is
disfavored. To summarize, from the fits of the Tables 2 and 3, plus the hints
from
data for the $q^2$ behavior of $f_+(q^2)$, among our assumptions of
extrapolation in mass
and $q^2$, the case (IV.2) seems to be favored, i.e., softened scaling plus
pole
behavior for $f_+(q^2)$ and constant behavior for $A_1(q^2)$.
	Lattice results of Table 4 have used the European Lattice Collaboration form
factors
\cite{21}. The $1/m_Q$ corrections to the asymptotic scaling law (28) were
fitted from the
lattice calculation, and for the $q^2$ dependence of $A_1(q^2)$ a two parameter
fit (pole +
constant) has been performed. The $q^2$ dependence  used here differs from the
one in
\cite{21} where pole dominance was assumed for all the form factors. A detailed
discussion is made on this matter in \cite{10}.
	Notice that the dilution factors are remarkably stable taking into account the
uncertainties involved and the variety of theoretical schemes, except for the
GISW
model \cite{15}, that owing to the results of Table 1 presumably cannot apply
at large $q^2$.

	Let us finally briefly comment on the strong phase $\delta_s$ defined in
(7)-(8) which is
the strong phase difference (up to the conventional minus sign in (8)) between
the
amplitudes $M(f) \equiv M(\bar B^0_s\to f)$ and and $\bar M(f) \equiv
M(B^0_s\to f)$ i.e. for example for $|f >
=|K^-(\vec p)D_s^+(-\vec p)>$ the sign between the two diagrams of
Figs. 2a and 2b. This phase appears in the asymmetry (13) and it is a further
hadronic uncertainty in the determination of the angle $\gamma$. It is clear
that
rescattering effects of the type $K^-D_s^+\to I\to K^-D_s^+$ where $I$ is some
intermediate state like $I = D^0\eta$ ,
... can induce strong phases in the amplitudes $M$ and $\bar M$. However, these
rescattering
effects are common to both amplitudes and cannot induce a strong phase shift
between
them. We can also have first a state with the quantum numbers of $c\bar u$
produced by the
weak interaction $\bar B^0_s, B^0_s \to I\to K^-D_s^+$ (with $I = D^0$,
$D^0\pi^0$, ... ). However, any strong
phase coming from the absorptive part of these contributions is also common to
the
amplitudes $M$ and $\bar M$, and does not introduce any strong phase shift. The
only
difference comes from the fact that the initial states are particle or
antiparticle,
and one could have for example intermediate states, produced by the strong
interaction, of the form  $\bar B^0_s\to B^+_uK^-$,... $ B^0_s\to B^-_uK^+$,
... followed by the
weak interaction. However, these intermediate states do not contribute to the
absorptive part of the amplitude, since the strong interaction conserves
flavor, and
these are not allowed decay products. Furthermore, these amplitudes give
equal
contributions to both processes   $\bar B^0_s\to B^+_uK^-$ and $ B^0_s\to
B^-_uK^+$ and cannot induce,
combined with absorptive processes, a difference of phase. Maybe a possible
source of
a strong phase shift could come from long distance strong effects that cross
the weak
interaction from the initial to the final state (short distance effects are
suposedly taken
into account by the effective color factor $a_1$). It seems to us that the
strong phase
$\delta_s$ is either small or it could even be zero, although the matter
deserves further investigation.

	In conclusion, we find that, summing over all decay modes with ground state
mesons
in the final state could be very useful for the determination of the CP angle
$\gamma$ of
the unitarity triangle, since most of the modes contribute with the same sign
to the
CP asymmetry. For the sum of the modes $K^-D^+_s$, $K^{\ast -}D^+_s$,
$K^-D^{\ast +}_s$, $K^{\ast -}D^{\ast +}_s$ we get a dilution factor $D\geq
0.6$, of the same
order of magnitude as for $K^-D^+_s$ alone, with a statistical gain
of the order of a factor 6. The dilution factor could be even larger owing to
the
favored extrapolation procedure from $D$ semileptonic form factors: $D_{tot}
\simeq 0.74$ for
softened scaling with a pole for $f_+(q^2)$ and roughly a constant for
$A_1(q^2)$. Taking
into account the variety of the theoretical schemes, the estimation of the
dilution
factor $D$ is surprisingly stable, although one must wait for a more precise
knowledge
of the heavy-to-light meson form factors to have a firm conclusion. The same
exercise
can be done for other sums over modes that can be useful in the determination
of the
angles $\alpha$ ($\pi\pi$, $\rho\pi$, $\rho\rho$) or $\beta$ ($\psi K$, $\psi
K^\ast$, ...).

\noindent {\bf Acknowledgements} \par  This work has been supported in part by
the
CEC Science Project SCI-CT91-0729 and by the Human Capital and Mobility Program
CHRX-CT93-0132. We aknowledge Patricia Ball and Daryl Scora for providing their
results for the heavy-to-light form factors, as well as the European Lattice
Collaboration for the numbers on heavy-to-light and heavy-to-heavy form
factors. One
of us (L. Oliver) acknowledges K. Goulianos and G. Apollinari for communicating
the CDF
collaboration data for the ratio $R_L$.

\vfill\supereject \centerline{\bf Figure Captions} \bigskip
\begin{enumerate}
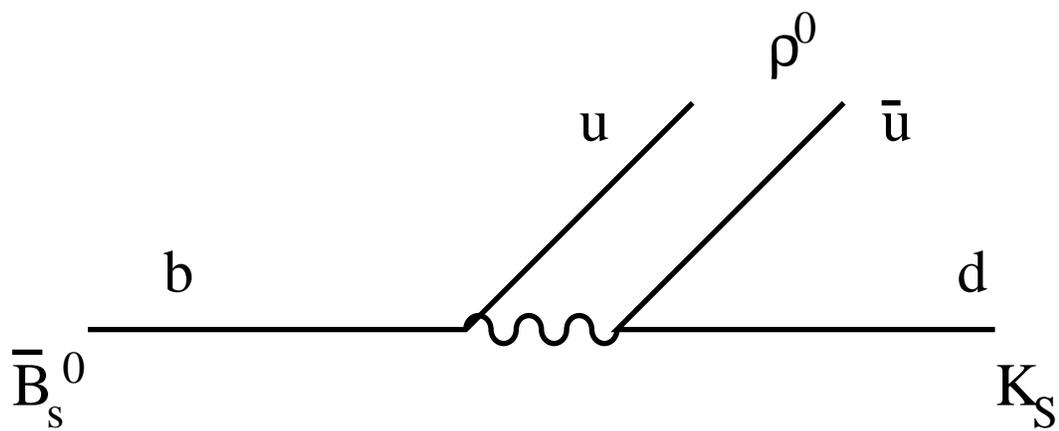
\item{\bf Fig. 1} The CP eigenstate mode $\bar{B}_s \to \rho^0K_S$.

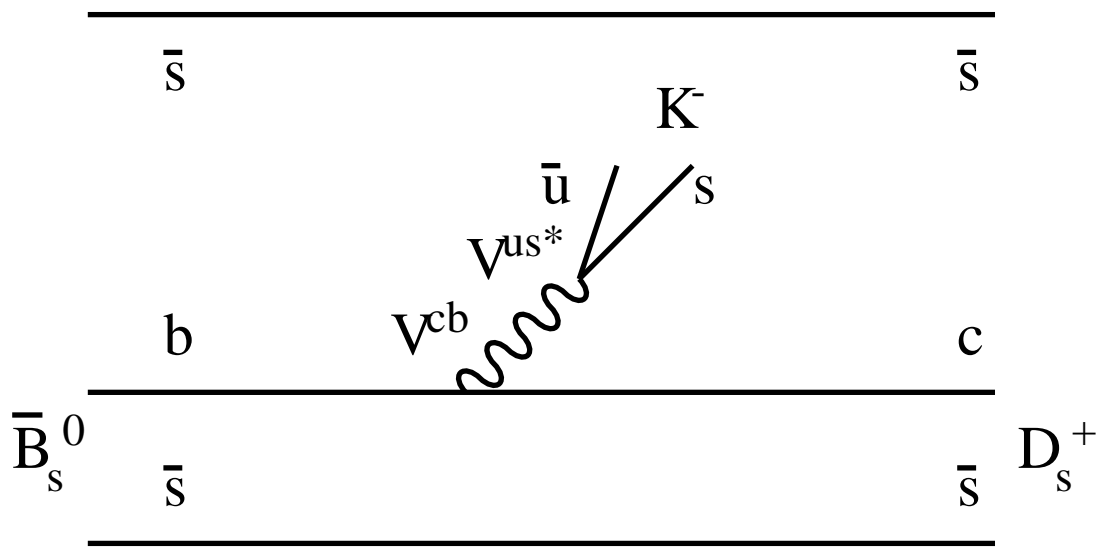
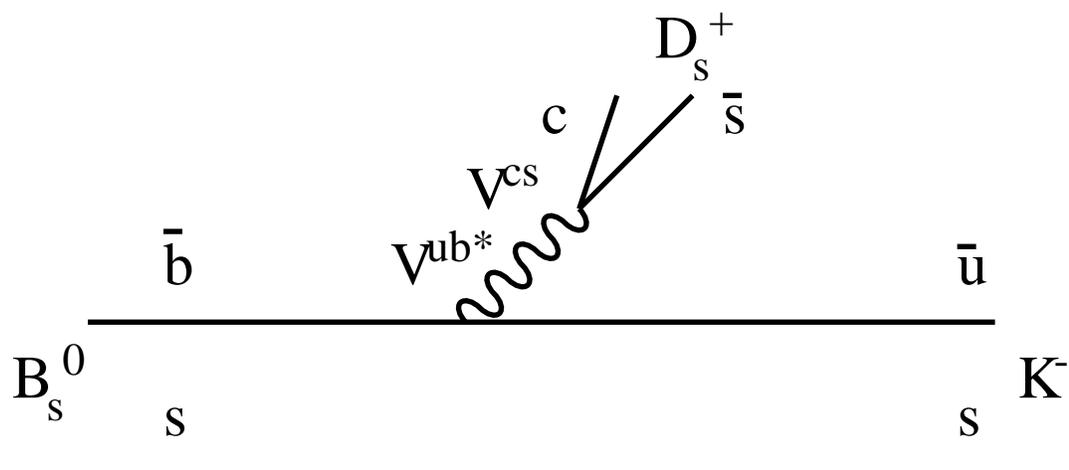
\item{\bf Fig. 2} The two spectator diagram contributions to the decay
$\bar{B}_s \to
K^-D^+_s$.
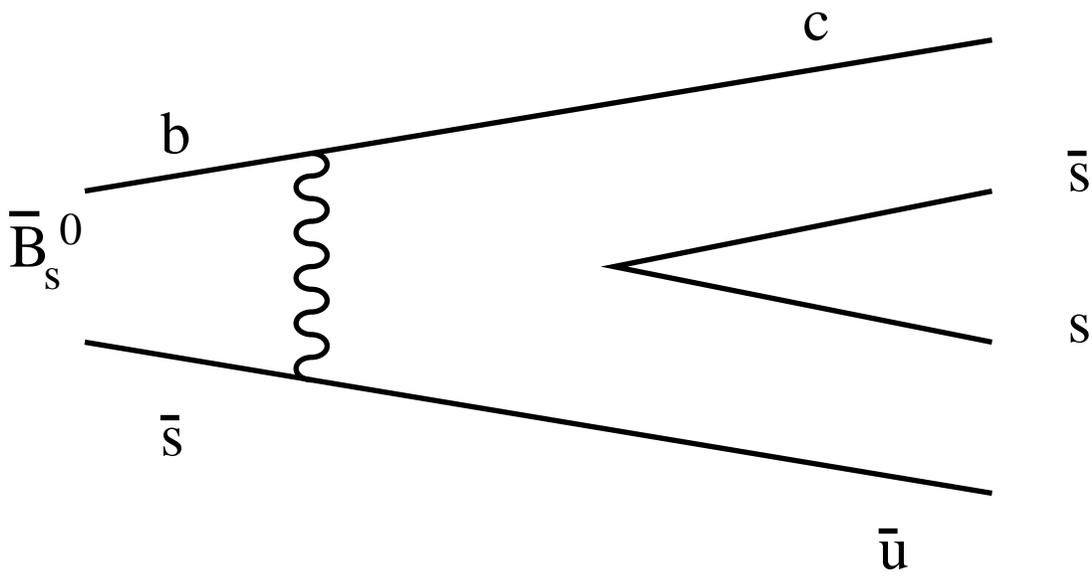
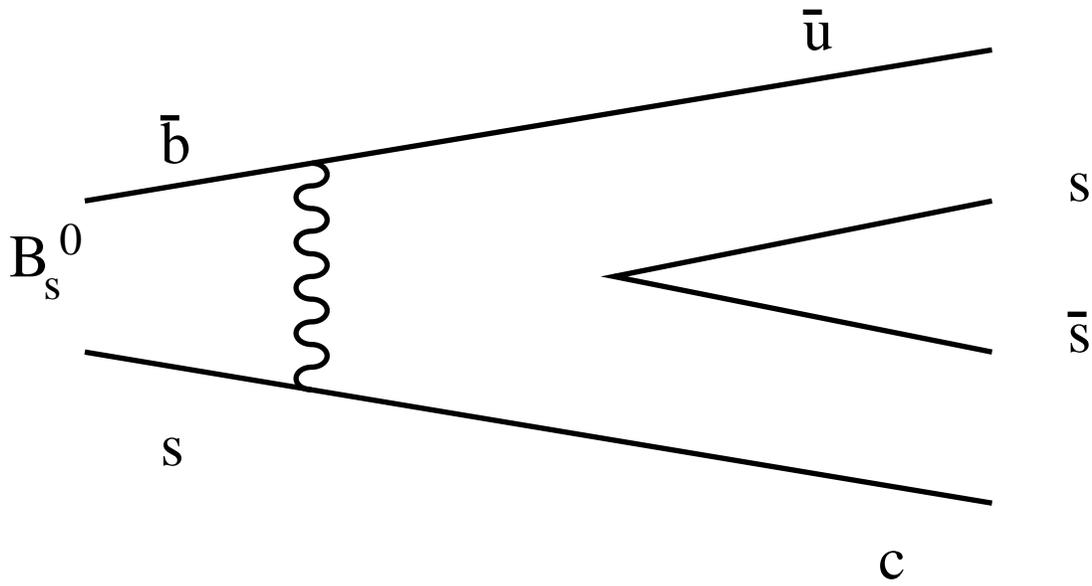
 \item{\bf Fig. 3}  Exchange diagram contributions to the decay $\bar{B}_s \to
K^-D^+_s$.
\end{enumerate}

\begin{thebibliography}{99}
 \bibitem{1} L. Wolfenstein, Phys. Rev. Lett. $\underline{51}$, 1945 (1983).
\bibitem{2} See, for a review, I.I. Bigi, V.A. Khoze, N.G. Uraltsev and A.I.
Sanda,
in~: CP Violation, ed. C. Jarlskog (World Scientific, Singapore, 1988). For a
more
recent review on the unitarity triangle, see Y. Nir, The CKM matrix and CP
violation,
Lectures presented at the Theoretical Advanced Study Institute on Elementary
Particle
Physics, Boulder, Colorado (1991), SLAC-PUB-5676. See also the recent nice
discussion
by I.I. Bigi, CERN-TH.7207/94, Invited Lecture given at the VIII Rencontres de
Physique de la Vall\'ee d'Aoste, La Thuile, March 1994.
\bibitem{3} R. Aleksan, I. Dunietz, B. Kayser and F. Le Diberder, Nucl. Phys.
$\underline{B361}$, 141 (1991).
\bibitem{4} R. Aleksan, A. Le Yaouanc, L. Oliver, O. P\`ene and J.-C. Raynal,
Phys.
Lett. $\underline{B317}$, 173 (1993).
\bibitem{5} J. Bernabeu and C. Jarlskog (DESY report 92-167) consider
the semi-inclusive modes $B_d$, $\bar{B}_d \to K_SX(c\bar{c})$ for which they
show
that they have the nice properties of pure CP eigenstates (dilution factor
equal to
one and absence of hadronic uncertainties in the asymmetry when summing over
$f$ and
$\bar{f}$. However, it is not clear how these semi-inclusive modes could be
separated
from modes of the type for example $B_d$, $\bar{B}_d \to (\pi^0K_S)_{K^*}
X(c\bar{c})$.
\bibitem{6} R. Aleksan, I. Dunietz and B. Kayser, Zeit. Phys.
$\underline{C54}$, 653
(1992). See also: R. Aleksan, B. Kayser and D. London, "In Pursuit of Gamma"
and R.
Aleksan, B. Kayser and P. Sphicas, "Measurement of the Angle Gamma", in
Proceedings
of the Summer Workshop on B Physics at Hadron Accelerators, Snowmass, Colorado
(1993).  \bibitem{7} N. Isgur and M.B. Wise, Phys. Lett.
$\underline{232B}$, 113 (1989) and $\underline{237B}$, 527 (1990)~; M.
Neubert and V. Rieckert, Nucl. Phys. $\underline{B382}$, 97 (1992).
\bibitem{8} N. Isgur and M. Wise, Phys.
Rev. $\underline{D42}$, 2388 (1990).
\bibitem{9} E.A. Paschos and R.A. Zacher, Zeit. f\"ur Phys. $\underline{C28}$,
521
(1985).
\bibitem{10}  R. Aleksan, A. Le Yaouanc, L. Oliver, O. P\`ene and J.-C. Raynal,
Orsay
preprint LPTHE 94/15 and Contribution to Beauty 94 Workshop, Mont Saint Michel,
France (April 1994), to be published in Nucl. Inst. and Meth. A (presented by
A. Le
Yaouanc), LPTHE 94/53.
\bibitem{11} M. Gourdin, A.N. Kamal and X.Y. Pham, Paris preprint
PAR/LPTHE/94-19 (1994).
\bibitem{12}. M. Wirbel, B. Stech and M. Bauer, Z. Phys.
$\underline{C29}$, 637 (1985)~; M. Bauer, B. Stech and M. Wirbel, Z. Phys.
$\underline{C34}$, 103 (1987).
\bibitem{13} G. Kramer and W.F. Palmer, Phys. Rev. $\underline{D45}$, 193
(1992). These authors make an extensive computation of $B \to VV$ modes, use
the
parametrization of ref. 12 for the form factors with the pole Ansatz. Their
result
corresponds to BSWI in Table 1, that gives a too large value for $R$.
\bibitem{14} M. Neubert,
V. Rieckert, B. Stech and Q.P. Xu, in Heavy Flavours, eds. A.J. Buras and M.
Lindner,
World Scientific, Singapore (1992).
\bibitem{15} N. Isgur et al. Phys. Rev. $\underline{D39}$, 799 (1989).
\bibitem{16} P. Ball, private communication, Phys. Rev. $\underline{D48}$, 3190
(1993).
\bibitem{17} M. Danilov, Talk given at the ECFA Working Group on $B$ Physics,
DESY
(1992).
\bibitem{18}  G. Apollinari and K. Goulianos, private communication. Results on
$B \to
\psi K(K^*)$ of the CDF collaboration have been presented at the 16th Int.
Symposium
on Lepton and Photon Interactions, Ithaca, N.Y. (1993), F. Abe et al.
Fermilab-Conf-93-200-E.
\bibitem{19} M.S. Witherell, Invited talk given at the
International Symposium on Lepton and Photon Interactions at High Energies,
Cornell
University, Ithaca, N.Y. (1993), UCSB-HEP-93-06.
\bibitem{20} A. Le Yaouanc et al., to
appear, LPTHE 94/64.
\bibitem{21} As. Abada et al. Nucl. Phys. $\underline{B374}$, 263 (1992);
$\underline{B376}$, 172 (1992).
\end{thebibliography}
\end{document}